\documentclass{elsart}

\usepackage{amssymb,amsfonts,graphicx}

\begin{document}

\begin{frontmatter}



\title{Characteristics of the Korean stock market correlations}


\author{Woo-Sung Jung\corauthref{cor1}}
\ead{wsjung@kaist.ac.kr} \corauth[cor1]{Corresponding author. Fax:
+82-42-869-2510.}
\author{, Seungbyung Chae, }
\author{Jae-Suk Yang, }
\author{Hie-Tae Moon}

\address{Department of Physics, Korea Advanced Institute of Science and Technology,
Daejeon 305-701, Republic of Korea}

\begin{abstract}

We establish in this study a network structure of the Korean stock
market, one of the emerging markets, with its minimum spanning tree
through the correlation matrix. Based on this analysis, it is found
that the Korean stock market does not form the clusters of the
business sectors or of the industry categories. When the MSCI
(Morgan Stanley Capital International Inc.) index is exploited, we
find that the clusters of the Korean stock market is formed. This
finding implicates that the Korean market, in this context, is
characteristically different from the mature markets.

\end{abstract}

\begin{keyword}
Correlation-based clustering \sep Emerging market \sep
Minimum spanning tree\sep Econophysics
\\

\PACS 89.65.Gh \sep 89.75.Fb \sep 89.75.Hc
\end{keyword}

\end{frontmatter}


\section{Introduction}
The stock price of a given company is a mutual inference of various
information, such as company revenue, competition performance,
currency policy, business barometers, political situation, and so
on. In other words, when the price is estimated, there are numerous
complicated factors that must be considered. In the stock market,
all companies are interconnected and consequently their stock prices
are correlated. This correlation, known as the potential of deep
inner impact, forms the stock market network.

Network theory has been extended into a wide range of
subjects\cite{faloutsos,ba-www,ba-bio,jung}. Barab\'asi and Albert
(BA) introduced the scale-free network which is constructed by the
growth rule and the preferential attachment rule\cite{ba}. We
consider the preferential attachment rule as the connectivity of an
influential company in stock market - a more influential company has
more connections with other companies. The interaction strengths
between nodes are important in many network systems. Non-binary
scale-free network\cite{wsf} which takes a continuous weight between
0 and 1 is a proper choice for modeling a stock market. We regard
companies as nodes (vertices) of the network, interacting relations
between stocks as links (edges) and correlation coefficients as
weights.

The minimum spanning tree (MST) is widely used to study the stock
market since Mantegna first constructed the network based on the
correlations\cite{mantegna}. The minimum spanning tree is generated
by selecting the most important links. We construct a correlation
matrix of N stocks. This matrix is symmetric and diagonal with
$\rho_{ii}=1$. The MST is determined by the distance matrix $D$
where $d_{ij}=\sqrt{2(1-\rho_{ij})}$. There have been several
attempts to identify the cluster
structure\cite{cluster1,cluster2,cluster3}. The MST is very useful
to observe the network topology and identify clusters of the market
including the stock and Foreign Exchange (FX)
market\cite{mantegna-mst,johnson}. Bonanno \emph{et al.} introduced
the topological properties of the MSTs through the real and model
markets' dataset\cite{bonanno}. Onnela \emph{et al.} investigated
the dynamical properties of the American market correlations and
taxonomy analysis in detail. The S\&P500 forms clusters with the
business sectors and the portfolio optimization with these clusters
is successful. The MST also can be applied to the portfolio analysis
in practice\cite{onnela}.

While there has been an abundance of literature concerning mature
markets - especially, the US market - relatively little work has
been published for emerging markets such as those of Korea, BRICs
and Eastern Europe. Emerging markets often lack liquidity and
reliable data, so they are generally unstable. These factors make
the study of emerging markets more complex. Even the universal
features for mature markets cannot be extended to emerging markets
for every cases\cite{india}. It seems that the model appropriate to
emerging market should be exploited.

In this paper, we aim to explore the topological characteristics of
the Korean market as a representative emerging market. We construct
the non-binary network by following the method introduced and
applied for S\&P500 companies by Kim \emph{et al}\cite{kim}. We
study the taxonomy and network topology of the Korean market with
it.

\section{Properties of the Korean Stock Market}
The Korean stock market is much smaller than the US stock market.
There are two stock markets in Korea - the Korea Stock Exchange
(KSE) and the KOSDAQ. There are 700 and 900 listed companies and
total capitalizations are \$400 billion and \$30 billion,
respectively. For NYSE and NASDAQ, there are thousands of listed
companies and the total market capitalization is approximately \$11
trillion. Rough estimation shows the US companies' average market
capitalization is ten times greater than that of Korea, which is
about \$2 million.

There are two predominant types of stock price indices; a
price-equally-weighted index is an arithmetic average and a
value-weighted index is an value weighted average of current stock
prices. The former such as DJIA assumes that every stock has the
same influencing power, while the latter such as S\&P500 assumes
that each stocks have the power proportional to their market
capitalization. KOSPI200 is a value-weighted index of 200
representative stocks in KSE, that is appropriate for the purpose of
this study.

Table \ref{table-top10} shows the market capitalization of some
largest companies listed on the S\&P500 and KOSPI200. This quantity
reflects the financial scale of a given company and the fraction of
total market capitalization in a stock market. ($\cdot$) denotes
stock symbol. The main distinction between two markets is the
influence of several top companies. The portion of top 10 KOSPI200
companies is 49.13\% where that of S\&P500 is 22.82\%. It means that
KOSPI200 index is more sensitive to the influence of a few top
companies.

Table \ref{table-investor} shows shareholding composition of KSE by
investor group. We can notice that foreign investors possess the
bulk of stocks of large companies and a large part of market
liquidity is supplied by them. Thus, the foreign investors' trading
activities are crucial to the KSE. So naturally the Korean stock
market is highly correlated with the foreign stock markets such as
the US market.

\section{Methodology}
We construct a network of KOSPI 200 companies; each node (company)
has a different number of links (connections) and weights
(correlations). We use the cross-correlations in stock price changes
between KOSPI200 companies from Jan/2001 to Jun/2004. The dataset is
daily closure prices in terms of Korean Won (KRW), the Korean local
currency. In 1997, the Asian financial crisis made the exchange rate
to fluctuate violently. However, during the period analyzed the
Korean FX market doesn't have an experience of external impact or
violent fluctuation. Let $Y_{i}(t)$ be the stock price of company
$i$. Then, the log-return of the stock price is defined as
\begin{equation}
S_{i}(t)=\ln Y_{i}(t+\Delta t)-\ln Y_{i}(t),
\end{equation}
where $\Delta$t is time interval. In this paper, we analyzed daily
data; $\Delta$t=1day. The cross-correlations between stock i and j
can be written as
\begin{equation}
\rho_{ij} =\frac{<S_{i}S_{j}>-<S_{i}><S_{j}>}
{\sqrt{(<S_{i}^{2}>-<S_{i}>^{2})(<S_{j}^{2}>-<S_{j}>^{2}) }}
\end{equation}
where $<\ldots>$ means a time average over the period. These
correlation coefficients form a correlation matrix $C$. This matrix
is a symmetric N$\times$N matrix. If stock i and j are completely
correlated (anti-correlated), $\rho_{ij}=+1(-1)$. The case of
$\rho_{ij}=0$ means they are uncorrelated.

Each node of the network corresponds to a company, which is fully
connected to every another nodes. Each link has a weight
$w_{ij}(=w_{ji})$, simply as the same value with the
cross-correlation coefficient; $w_{ij}=\rho_{ij}$.

The influence strength (IS) is a physical quantity to measure how
strongly a node influences other nodes. This quantity is defined as
the sum of the weights of all links incident upon a given
node $i$,
\begin{equation}
q_{i}=\sum _{j\ne i} w_{ij},
\end{equation}
where $j$ denotes the links connected to the node $i$. Since the
weight is distributed in the range [-1,1], the IS can be negative.
Here we just need to determine the influencing structure, and the
sign of $q_{i}$ is not important. Thus we only deal with the
absolute magnitude of the IS denoted by $|q_{i}|$.

\section{Characteristics of the Korean Stock Market}
\begin{figure}
\includegraphics[angle=0,width=1.0\textwidth]{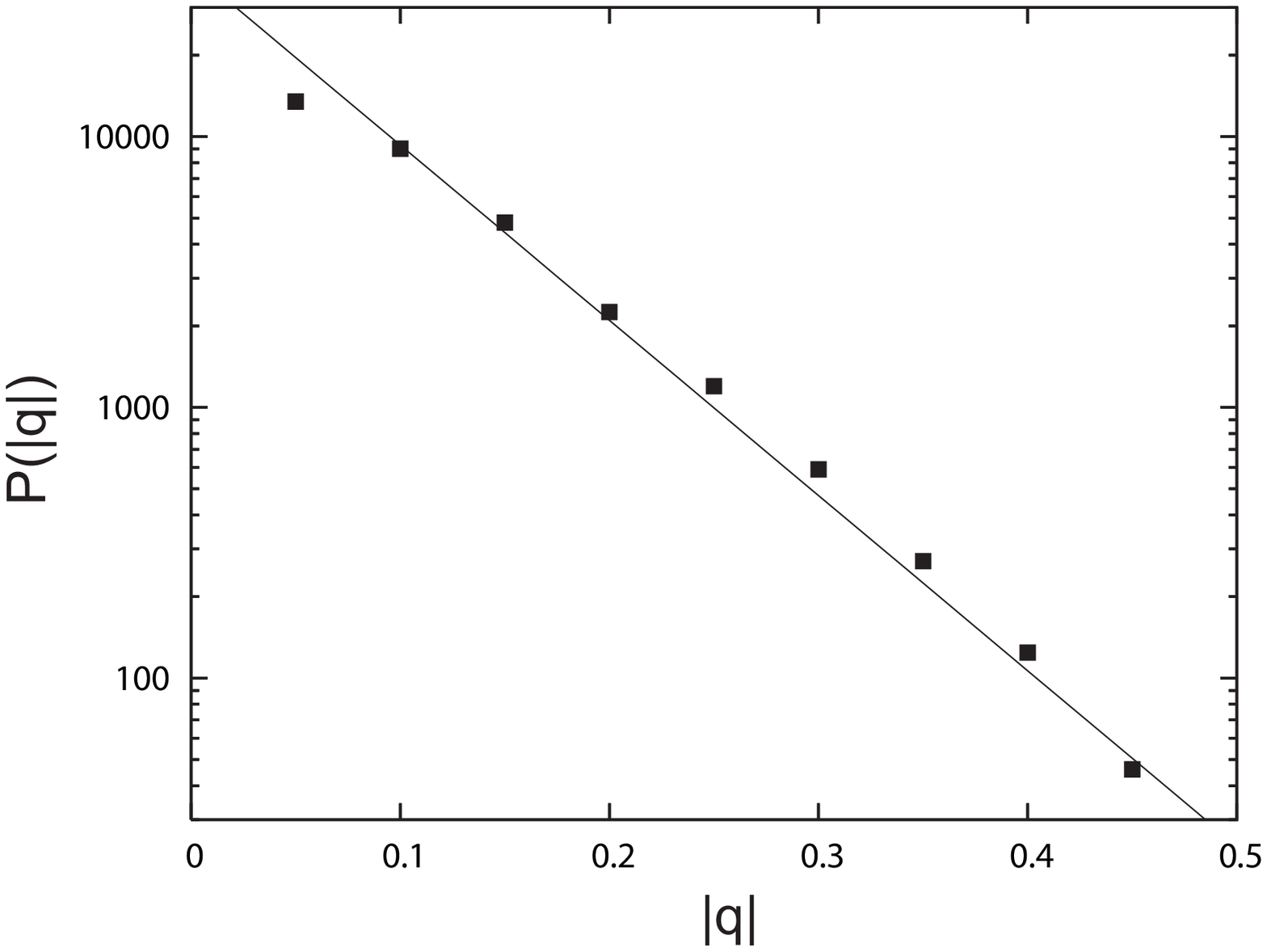}
\caption{Plot of the KOSPI 200's influence strength (IS)
distribution versus the absolute magnitude of the
influence strength. The slope of the guide line is 6.5
}
\label{kospipowerlaw}
\end{figure}

In Fig. \ref{kospipowerlaw}, we plot the IS distribution
$P_{K}(|q|)$ of KOSPI200. Kim \emph{et al}.\cite{kim} found the IS
distribution of S\&P500 follows a power law distribution,
$P_{SP}(|q|)\sim |q|^{-\eta}$ where the exponent $\eta$ is estimated
to be $\eta_{SP}\sim 1.8$. It is known that as the degree exponent
is smaller in SF networks, the connectivity of a node with a large
degree becomes higher, and hence the network tends to be more
centralized to a few nodes. In other words, several powerful
companies make a dominant effect to the whole market. It is also
noticeable that the IS component of the S\&P500 is smaller than the
degree exponent values for SF networks in the real world such as the
Internet and the World-Wide Web\cite{faloutsos,ba-www}.

By Table \ref{table-top10}, the largest company from the viewpoint
of a market capitalization in the S\&P500 is General Electric (GE);
its fraction is 3.39\%. In the case of the KOSPI200, Samsung
Electronics Corporations (SEC) occupies this position; its fraction
is 21.94\%. So we can consider the KOSPI200's hub to be more
powerful, and the KSE network is more centralized to this company
than the S\&P500. However, Fig. \ref{kospipowerlaw} shows that the
KOSPI200's IS distribution does not follow a power law distribution,
but an exponential distribution. The Korean stock market is less
centralized than the S\&P500 or other scale-free networks in the
real world.

\begin{figure}
\includegraphics[angle=0,width=1.0\textwidth]{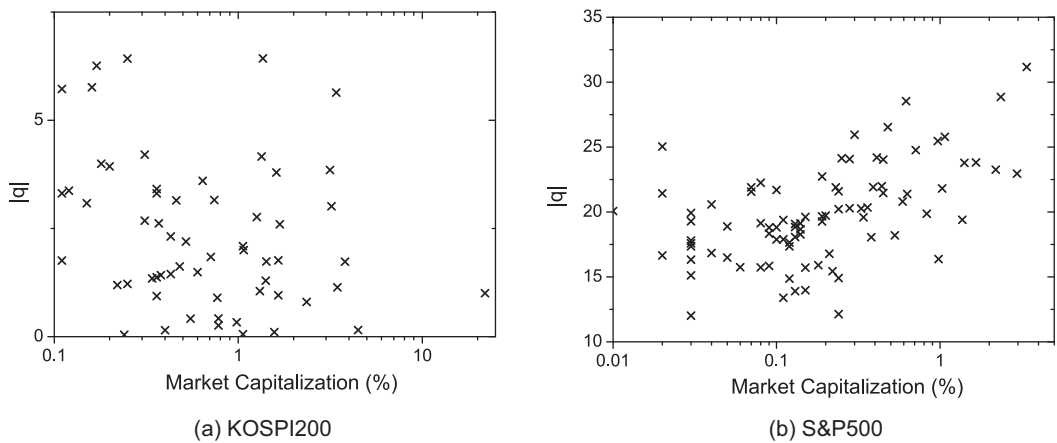}
\caption{Positive correlations between market capitalization
and $|q|$ are appeared in the S\&P500.
But the KOSPI200 has no correlations.}
\label{isdist}
\end{figure}

We construct the asset tree through the minimum spanning tree (MST)
to find the difference between the S\&P500 and KOSPI200. GE acts as
the hub of the S\&P500's MST\cite{mantegna,onnela}. However, we
cannot find any comparable hub in the whole market of the KOSPI200
contrary to the dominant position of SEC in KSE. SEC node is located
far away from the center (Fig. \ref{isdist}). This means the
fraction of SEC's market capitalization is large, and as such the
KOSPI Index moves with SEC while most companies' stocks do not
follow this trend. Neither SEC nor the others in Table
\ref{table-top10} is located at the center of KOSPI's MST. The weak
influence of SEC shows weak correlations with other stocks. Hence,
we cannot find such scale-free behaviors in the Korean market as the
American market.

One possible explanation of this difference is the market maturity.
The American stock market is a mature market. There are numerous
powerful companies such as GE, Microsoft and Citigroup. These
companies have similar market capitalization and influence power on
the market. In the Korean stock market, an emerging market, there is
a great gap between the SEC and the others - even though, 9
companies of Table \ref{table-top10} - with a viewpoint of the
market capitalization.

For the application of portfolio optimization, the identification of
groups of stocks in common dynamics is necessary to diversify the
risks. At first, we introduce some terminology. The term
\emph{branch} is defined as a subset of a tree, to all nodes that
share a specified common parent and \emph{cluster} as a subset of a
branch. There are two kinds of clusters. One is a \emph{complete}
cluster and the other is \emph{incomplete}. A complete cluster
contains all the companies of the studied set belonging to the
corresponding branch, so that none are left outside the cluster.
Onnela \emph{et al}.\cite{onnela} found that clusters of S\&P500
with business sector or industry categories are mostly incomplete,
but come very close to being complete clusters, only missing one or
two companies of the cluster. We consider this situation as a
complete cluster from the viewpoint of practical portfolio
optimization. However, the clusters of KOSPI200 do not coincide with
business sectors or industry categories. In addition, they made the
portfolio using the central node (GE), but the KOSPI200 has no
single central hub, and thus this method cannot be applied to the
Korean situation.

We attempted to identify groups of KOSPI200 with other rules than
business sectors or industry category. Most major Korean companies
are members of conglomerate forms of enterprise, commonly known as
\emph{Chaebeol}. For example, Samsung Electronics Co. is a member of
Samsung Conglomerate. This conglomerate is comprised of many
companies over different sectors, i.e., Samsung Electronics Co.,
Samsung Life Insurance, Samsung Heavy Industries Co., Samsung
Petrochemical Co., Samsung Corporation, and so on. Their ownerships
are controlled by complex shareholding structures. So their stocks
can be considered a group. However, we cannot find any group
structure related with Korean major conglomerates - Samsung, LG, SK,
Hyundai, and so on.

We also consider the influence of foreign investors on the trading
patterns of domestic investors. Foreign investors are generally
believed to employ superior techniques and information and their
strategies are considered as benchmarks by domestic counter parties
in the Korean stock market. So we apply the MSCI index to make
groups of Korean stocks. Morgan Stanley Capital International Inc.
(MSCI) is one of the leading providers of equity indices and offers
the most widely used international equity benchmarks by
international investors. MSCI Equity Indices are designed to fulfill
the investment needs of a wide variety of global institutional
market participants. These include many categories of indices, i.e.
Sector, Industry Group and Industry Indices, Global, Regional and
Country Equity indices, and so on. We focus on the MSCI Korea Index
- one of the MSCI Country Equity Indices.

\begin{figure}
\includegraphics[angle=0,width=1.0\textwidth]{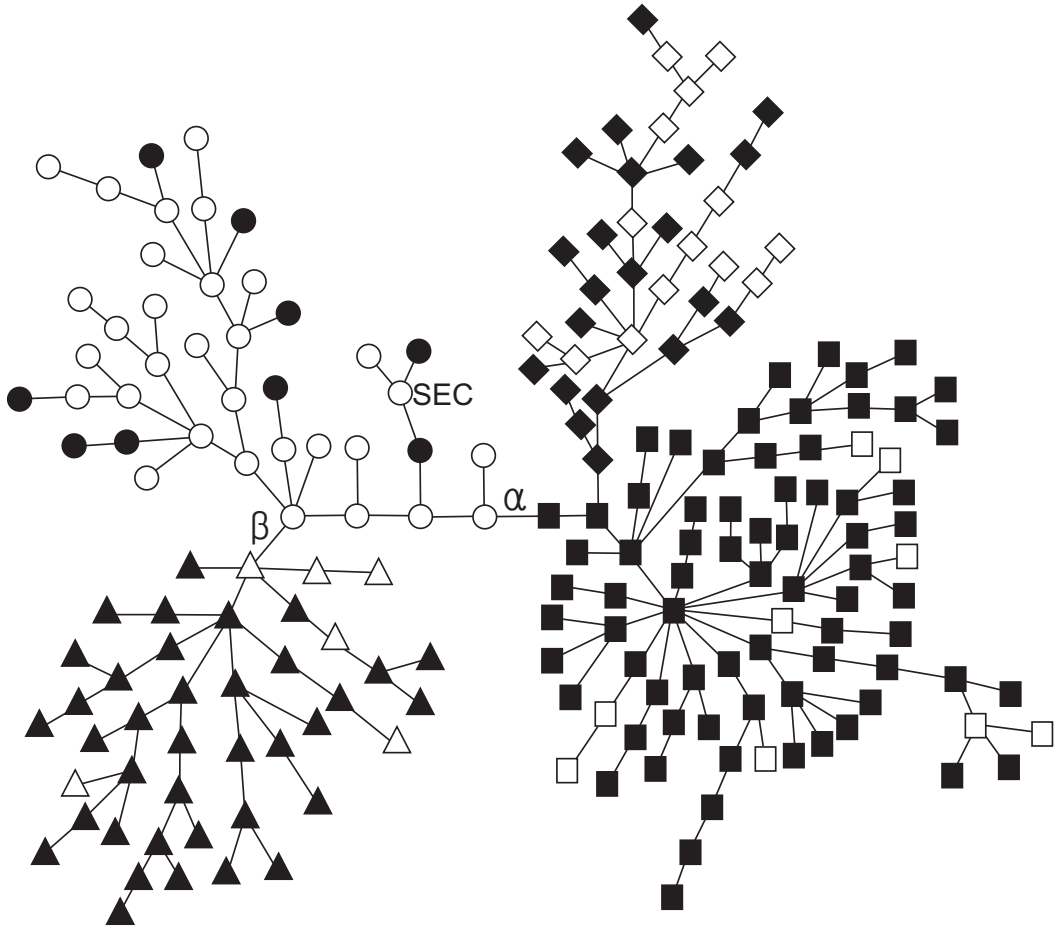}
\caption{A minimum spanning tree of the KOSPI 200. White spots mean
the stocks included in the MSCI index, black spots mean the stocks
which are not included. There are 4 clusters - $\mathcal{A}$:
rectangle ($\square$, $\blacksquare$), $\mathcal{B}$: circle
($\circ$, $\bullet$), $\mathcal{C}$: triangle ($\vartriangle$,
$\blacktriangle$) and $\mathcal{D}$: diamond ($\lozenge$,
$\blacklozenge$). $\mathcal{A}$ and $\mathcal{C}$ are the clusters
of the stocks which are not listed on the MSCI index, $\mathcal{B}$
is the cluster of the stock listed on the MSCI index. $\mathcal{D}$
is shuffled area, so cannot be a meaningful cluster. } \label{mst}
\end{figure}

Fig. \ref{mst} supports the validity of MSCI index grouping. We can
identify two types of clusters by mainly composed of stocks included
in MSCI index or not. While all of them are incomplete clusters,
they can be considered as complete clusters in practice. It seems
that MSCI index grouping is the most acceptable method for the
Korean market. In fig \ref{mst}, the weight of link $\alpha$ is the
second lowest and that of $\beta$ is the fifth lowest one. So we can
divide the whole market into three clusters separated by these two
links, and each cluster forms a sub-market. Another noticeable
feature of this MST is the absence of a global hub. The node in the
center of cluster seems to be a hub, but it is just a local hub for
isolated sub-market. We can also see that highly capitalized stock
like SEC is far from a hub even for a sub-market $\mathcal{B}$.

Bonanno \emph{et al.}\cite{bonanno} constructed the MST using market
models; random market model and one-factor model. Random market
model assumes that the return distribution is uncorrelated Gaussian
and one-factor model assumes that the return is controlled by a
single factor like index. The degree distribution for the MST of
mature market follows a power law distribution, and the degree
distribution of the one-factor model is decayed rapidly and contains
an asset with a very high value of the degree. Fig.
\ref{kospipowerlaw} shows that the degree distribution of the
KOSPI200 follows neither a power law distribution nor a distribution
of the one-factor model. It seems close to a distribution of the
random market model.

\section{Conclusions}
We have studied the Korean stock market and obtained some
characteristics that differ from the characteristics of the US
market. The pertinent question is, why does the Korean stock market
have different properties? One possible reason is the composition of
firms. The history of mature markets is longer than that of emerging
markets. Thus, the mature markets have many companies including
several large firms. In the case of the Korean market, there are
only a few large firms, e.g. SEC; these corporations are very large
in comparison with others. As such, these large firms are separated
from other companies of the market. This accounts for why there are
no hubs in the Korean stock market. We don't know yet whether this
is the characteristics of an emerging market or only Korean
characteristics. The other is the trading culture and globalization.
Foreigners' trading patterns are much important in the Korean
market. Globalization has progressed very rapidly and influence of a
few developed countries has become more and more powerful. At
present, many stock markets' synchronization to the US market is
observed. In other words, the whole markets in the world are
synchronized. We may thus find clusters in terms of the MSCI index.
If a specified company's stock is included in the MSCI index, it is
more synchronized to a foreign market and regarded as a good
company's stock to the Korean market. All markets throughout the
world have characteristics of their own. We need to study each
market with its own properties.

The 1997 Asian financial crisis was a very important event to the
Korean market. After the crisis, the market's response to the
external market is more sensitive\cite{climent}. The correlation
coefficient of the Korean market is smaller than that of the
American market and sometimes shows unusual distribution. The
correlation and the MST have more information about the market than
this paper's analysis, i.e. average length, positive correlation and
negative correlation. The investigation about the points mentioned
with the knowledge on the history of the Korean market is our future
work.

We wish to thank S.-W. Son, O. Kwon and C. Kim
for helpful discussions and supports.





\newpage

\newpage
\clearpage

\begin{table}
\caption{10 top listed companies by market capitalization in
S\&P500(except MER and MWD) and KOSPI200.(Jun. 2004)}
\label{table-top10}
\begin{tabular}{|c|c|c|c|c|}
\hline
Rank&\multicolumn{2}{|c|}{S\&P 500}&\multicolumn{2}{|c|}{KOSPI 200}\\
\cline{2-5}
&Company(Symbol)&\%
&Company(Symbol)&\%\\
\hline
1&General Electric (GE) & 3.39 & Samsung Electronics (005930) & 21.94\\
\hline
2&Exxon Mobil (XOM) & 2.96 & SK Telecom (017670) & 4.48\\
\hline
3&Microsoft (MSFT) & 2.91 & POSCO (005490) & 3.80\\
\hline
4&Pfizer INC. (PFE) & 2.39 & Kookmin Bank (060000) & 3.46\\
\hline
5&CitiGroup (C) & 2.36 & KEPCO (015760) & 3.41\\
\hline
6&Wal-Mart (WMT) & 2.19 & Hyundai Motors (005380) & 3.21\\
\hline
7&Amer.Intl.Group (AIG) & 1.82 & KT (030200) & 3.15\\
\hline
8&Bank of America (BAC) & 1.73 & LG Electronics (066570) & 2.35\\
\hline
9&Johnson\&Johnson (JNJ)& 1.66 & SK Corp. (003600) & 1.68\\
\hline
10&P\&G (PG) & 1.41 & Woori Finance (053000) & 1.65\\
\hline
\end{tabular}
\end{table}

\begin{table}
\caption{Shareholding by investor group(2003)}
\label{table-investor}
\begin{tabular}{|c|c|c|c|}
\hline
&Individual&Foreigners&Institution and Others\\
\hline
\# of shareholder(A) & 99.33\% & 0.39\% & 0.22\% \\
\hline
\# of shares(B) & 48.50\% & 13.99\% & 37.51\% \\
\hline
Market capitalization(C) & 23.44\% & 37.67\% & 38.89\% \\
\hline
B/A & 0.488 & 35.8 & 170.5 \\
\hline
C/A & 0.236 & 96.6 & 176.8 \\
\hline
C/B & 0.483 & 2.70 & 1.04 \\
\hline
\end{tabular}
\end{table}

\end{document}